\documentclass[aps,prl,notitlepage,superscriptaddress,showpacs,twocolumn ]{revtex4-1}
\usepackage{graphicx,subfigure,epsfig}
 \usepackage{array,multirow}
\usepackage{dcolumn}
\usepackage{amssymb,amsmath,amsfonts,mathrsfs}
\usepackage{array}
\usepackage{times,setspace}
\usepackage{latexsym}
\usepackage{float,flafter,bm,bbm}
\usepackage{epstopdf,color,multirow}
\usepackage[colorlinks,linkcolor=blue,anchorcolor=blue,urlcolor=blue,citecolor=blue]{hyperref}
\usepackage{footnote}

\hypersetup{
    colorlinks=true,
    linkcolor=blue,
    filecolor=magenta,
    urlcolor=blue,
}

\begin{document}

\title{Superconductivity in doped triangular Mott insulators:\\
the roles of parent spin backgrounds and charge kinetic energy}

\author{Zheng Zhu}
 \email{zhuzheng@ucas.ac.cn}
\affiliation{Kavli Institute for Theoretical Sciences, University of Chinese Academy of Sciences, Beijing 100190, China}
\affiliation{CAS Center for Excellence in Topological Quantum Computation, University of Chinese Academy of Sciences, Beijing, 100190, China}
\author{Qianqian Chen}
\email{chenqianqian@ucas.ac.cn}
\affiliation{Kavli Institute for Theoretical Sciences, University of Chinese Academy of Sciences, Beijing 100190, China}

\begin{abstract}
We study the prerequisites for realizing superconductivity in doped triangular-lattice Mott insulators by considering three distinct parent spin backgrounds, i.e.,  $120^{\circ}$ antiferromagnets, quantum spin liquid, and stripy antiferromagnets, and all possible sign combinations $(\tau_1, \tau_2)$ of nearest-neighbor hopping and next-nearest-neighbor hopping $(t_1, t_2)$. Based on density-matrix renormalization group calculations, we find that, with finite $t_2$ and specific sign combinations $(\tau_1, \tau_2)$, the quasi-long-range superconductivity order can always be achieved, regardless of the nature of the parent spin backgrounds.
Besides specific hopping signs $(\tau_1, \tau_2)$, these superconductivity phases in triangular lattices are commonly characterized  by short-ranged spin correlations and two charges per stripe. In the robust superconductivity phase realized at larger $t_2/t_1$, flipping the signs $\tau_2$ and $\tau_1$ gives rise to the stripe phase without strong pairing and pseudogap-like phase without Cooper-pair phase coherence, respectively.
Interestingly, the roles of the two hopping signs are switched at smaller $t_2/t_1$.
Moreover, different sign combinations $(\tau_1, \tau_2)$ would stabilize distinct phases including superconductivity, charge density waves, spin density waves, and pseudogap-like phases accordingly. Our findings suggest the important role of charge kinetic energy in realizing superconductivity in doped triangular-lattice Mott insulators.
\end{abstract}

\maketitle

\emph {Introduction.---} Understanding the superconductivity  (SC) that emerges from doping Mott insulators has been a long-standing issue in physics.
Unlike the conventional Bardeen-Cooper-Schrieffer (BCS) superconductivity~\cite{BCS},
the prerequisites for achieving superconductivity in doped Mott insulators remain  elusive~\cite{Anderson1987,FCZhang1988,Dagotto1994,DNSheng1996,Lee06,XiaoGangWen1996, Balents2007,Keimer2015,Fradkin2015,Senthil2005, KWu2008,ZhengYuWeng2011,ZhengZhu2014,ZhengZhu2018,Arovas2022,MingpuQin2022}.
The general understanding
originates from doping quantum spin liquids (QSLs)~\cite{QSL0,QSL1,QSL2,Broholm2020,Knolle2019}, which is composed of condensed spin resonating-valence-bond pairs such that doped charges have an energetic incentive to pair~\cite{Anderson1987,Lee06,Keimer2015}. The superconductivity arises when those pairs achieve long-range phase coherence. However, besides QSL,  the parent spin backgrounds usually host various magnetic orders, then it is highly instructive to explore the doped distinct magnetic ordered Mott insulators to fully reveal the prerequisites for realizing superconductivity. Moreover, since the interplay between the doped charge and the spin backgrounds determines the charge properties~\cite{Dagotto1994,DNSheng1996,Lee06, KWu2008,ZhengZhu2014}, it is also fundamentally significant to identify the role of charge kinetic energy in the resulting superconductivity.

Tremendous efforts on this issue have  been devoted to square lattice ~\cite{Anderson1987,FCZhang1988,Lee06,Dagotto1994,Keimer2015,DNSheng1996,XiaoGangWen1996,Fradkin2015,Senthil2005,Balents2007, KWu2008,ZhengYuWeng2011,ZhengZhu2014, ZhengZhu2018,Arovas2022,MingpuQin2022} since the discovery of high-temperature superconductivity in cuprates.
Nevertheless, such studies on the triangular lattice have equal importance and the geometric frustrations bring even richer physics for both Mott insulators~\cite{Saarikoski2015,Shirakawa2017,Szasz2020,YaHuiZhang2021, BinBinChen2021, Wietek2021,Cookmeyer2021,Szasz2021, YiqingZhou2022} and doped Mott insulators ~\cite{Baskaran2003,Watanabe2004,Motrunich2004NaCoO1,Motrunich2004NaCoO2, QiangHuaWang2004, Kumar2003,SenZhou2008,Raghu2010,Kiesel2013,Zhu2020,Jiang1, Sheng1,WenYuHe2018,XueYangSong2021,MingpuQin2022IOP,KevinHuang2022,Chen2022, Jiang2, Sheng2,YangZhang2022,Wietek2022,XueYangSong2022}, as revealed  in earlier experimental discoveries of superconductivity in Na$_x$CoO$_2$$\cdot$yH$_2$O~\cite{NaCoO1,NaCoO2}. Remarkably, distinct hopping signs are inequivalent on triangular lattice~\cite{Chen2022,Zhu2020}.
More recently, both the cold-atom and condensed-matter experiments have offered additional platforms to probe the doped triangular-lattice Mott insulators with a wide range of parameters in a well-controlled manner, such as loading ultracold fermions onto triangular optical lattice~\cite{Yang2021,Bohrdt2021}, stacking two distinct transition-metal dichalcogenides (TMD) like  WSe$_2$/WS$_2$ hetero-bilayers~\cite{Tang2020,Xu2020,Kennes2021,FengchengWu2018},  depositing atomic layers on semiconductor substrates such as Sn/Si(111)~\cite{SnSi111} and doping organic compounds \cite{Oike2015,Oike2017}.
Remarkably, these platforms host different parent spin backgrounds and different sign combinations in charge hoppings~\cite{NaCoO1,NaCoO2,Oike2015,Oike2017,Tang2020,Xu2020,SnSi111,Kennes2021,FengchengWu2018}, laying the experimental foundations for studying the roles of  Mott insulation and kinetic energy in resulting superconductivity.

\begin{figure}[tp]
\begin{center}
\includegraphics[width=0.37\textwidth]{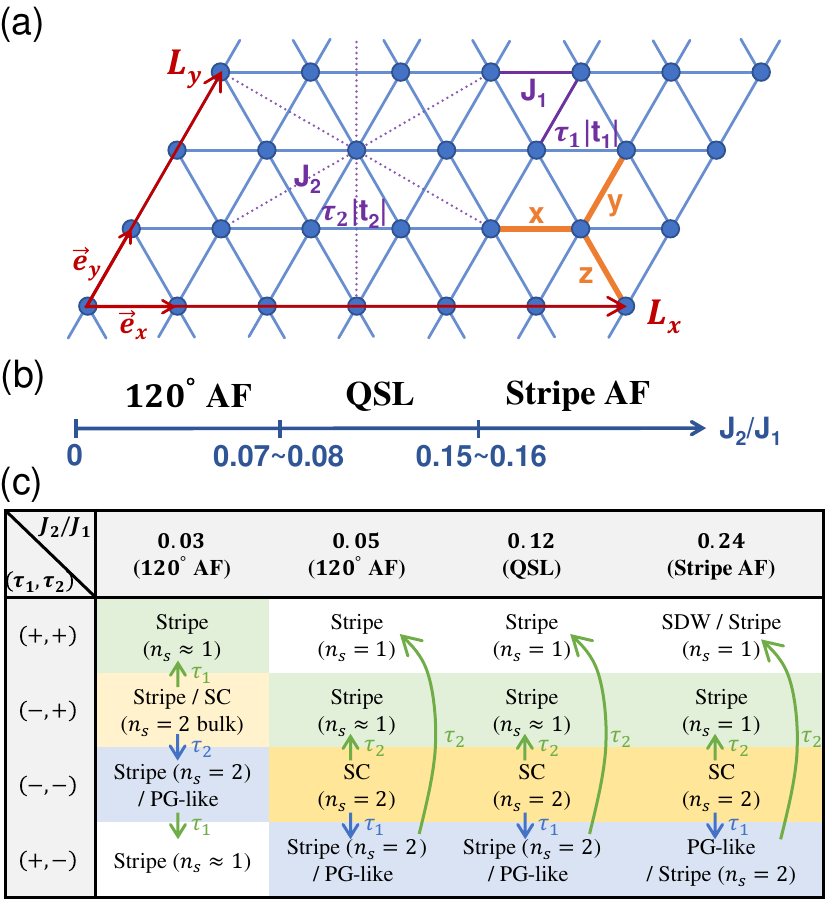}
\end{center}
\par
\renewcommand{\figurename}{Fig.}
\caption{(a) Sketch of triangular lattice and model parameters. (b) Phase diagram of $J_1$-$J_2$ Heisenberg model. (c) The most dominant correlations in lightly doped Mott insulators on YC4 cylinders controlled by  $J_2/J_1$=$(t_2/t_1)^2$  for different  hopping signs $(\tau_1, \tau_2)$.}
\label{Fig_TriangularLattice}
\end{figure}

\begin{figure*}[tbp]
\begin{center}
\includegraphics[width=0.88\textwidth]{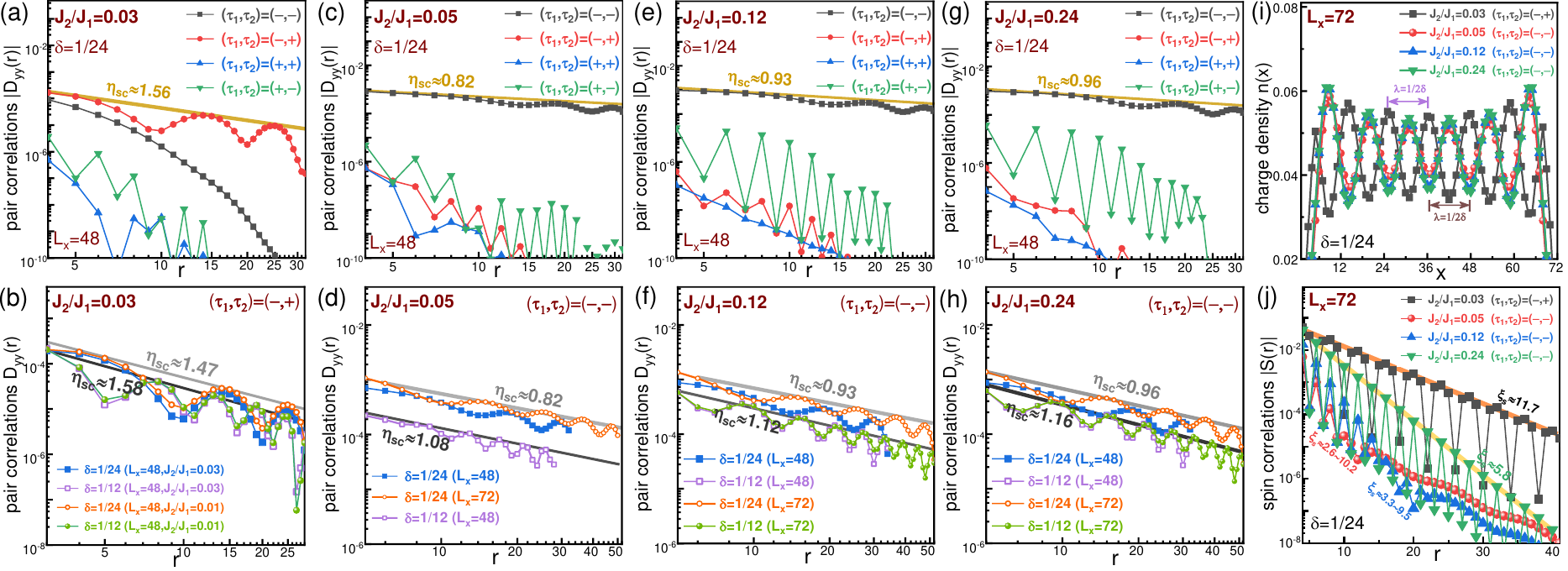}
\end{center}
\par
\renewcommand{\figurename}{Fig.}
\caption{
(a-h) The pair correlations for all hopping signs $(\tau_1, \tau_2)=(\pm,\pm)$ in lightly doped three  parent Mott insulating spin backgrounds: $120^{\circ}$ AF (a-d), QSL (e, f) and stripe AF (g, h).
Here, we consider systems with  $L_y=4$ and $\delta=1/24, 1/12$. (i,j) The charge density distribution (i) and the spin correlations (j) at the parameters with quasi-long-range superconductivity order.
}
\label{Fig_DD}
\end{figure*}

Motivated by the above, we examine the roles of the parent spin backgrounds and the charge hopping signs in achieving triangular-lattice superconductivity. We consider three distinct spin backgrounds realized in the $J_1$-$J_2$ model,
\begin{equation}
H_{\mathrm{J_1-J_2}}=J_1\sum_{\left\langle ij\right\rangle } \mathbf S_i\cdot \mathbf S_j+J_2\sum_{\left\langle\left\langle ij\right\rangle\right\rangle } \mathbf S_i\cdot \mathbf S_j,
\end{equation}
where $\langle ij\rangle $ and $\langle\langle ij\rangle\rangle$ denote the nearest-neighbor (NN) and next-nearest-neighbor (NNN) bonds, respectively.
The undoped spin backgrounds are $120^{\circ}$ antiferromagnetic (AF)  at $J_2/J_1\lesssim 0.07\sim0.08$, stripy AF at $J_2/J_1\gtrsim0.15\sim0.16$ and QSL in between  \cite{ FaWang2006,
ZhenyueZhu2015, WenJunHu2015, Saadatmand2016,Iqbal2016, Wietek2017,ZhenyueZhu2018,ShijieHu2019, Syromyatnikov2022, TaTang2022,Drescher2022, Sherman2022}, as shown in Fig.~\ref{Fig_TriangularLattice}(b). The motion of the doped charge can be captured by
 \begin{equation}\label{eq:model}
  \begin{aligned}
H_{\mathrm{t_1-t_2}}=\,&\tau_1|t_1|\mathcal{P}\sum_{\left\langle ij\right\rangle \sigma}(c_{i\sigma}^{\dagger}c_{j\sigma}+h.c.)\mathcal{P}\\
&+\tau_2|t_2|\mathcal{P}\sum_{\left\langle\left\langle ij\right\rangle\right\rangle \sigma}(c_{i\sigma}^{\dagger}c_{j\sigma}+h.c.)\mathcal{P},
 \end{aligned}
\end{equation}
where ${c_{i\sigma }^{\dagger }}$ is the fermion creation operator, and $\mathcal P$ projects to the single-occupancy subspace. $\tau_1=\pm$ ($\tau_2=\pm$) denote the signs of NN hopping $t_1$ (NNN hopping $t_2$).
The different sign combinations $(\tau_1, \tau_2)$ are physically reasonable because they correspond to different materials ~\cite{NaCoO1,NaCoO2,Oike2015,Oike2017,Tang2020,Xu2020,SnSi111,Kennes2021,FengchengWu2018}.
Physically, $J_2/J_1$=$(t_2/t_1)^2$ since \eqref{eq:model} is an effective Hamiltonian of the Hubbard model in the strong coupling limit.

\begin{figure*}[tp]
\begin{center}
\includegraphics[width=0.88\textwidth]{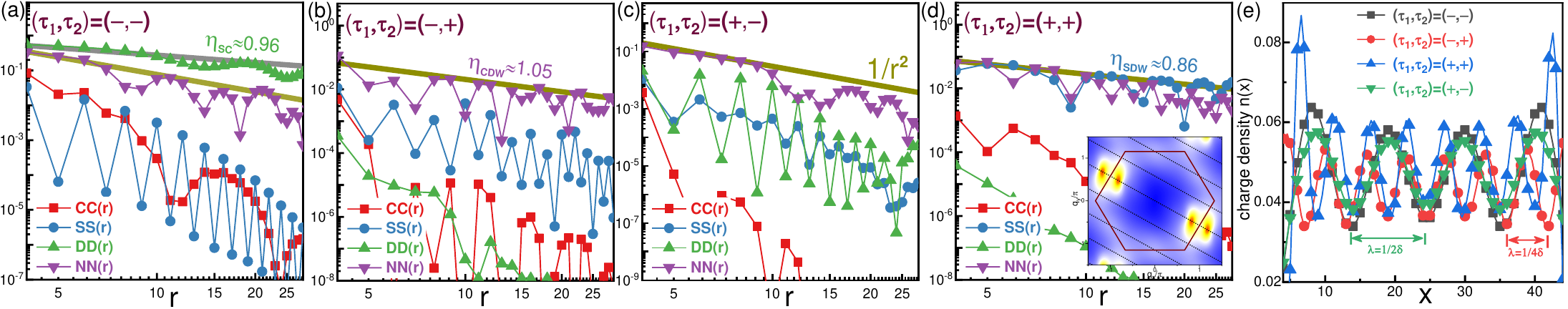}
\end{center}
\par
\renewcommand{\figurename}{Fig.}
\caption{Various correlations for  distinct hopping signs
$(\tau_1, \tau_2)=(\pm,\pm)$ (a-d) at lightly doped ($\delta=1/24$) stripe AF parent spin background.
(e) The charge density distributions for distinct $(\tau_1, \tau_2)$. Here,  we consider $J_2/J_1=0.24$ (stripe AF) on $N=48\times4$ cylinders. The doped QSL with $J_2/J_1=0.12$ and doped $120^{\circ}$ AF with $J_2/J_1=0.05$ exhibit similar behavior~\cite{SM}.}
\label{Fig_StripeSC}
\end{figure*}

We study the ground-state properties by density matrix renormalization group (DMRG)~\cite{DMRG1,*DMRG3,DMRG2}.
To examine the roles of Mott insulation after doping, we focus on light doping and study three distinct Mott insulators by tuning $J_2/J_1$.
The  triangular lattice is spanned by the primitive vectors  $\mathbf{e_x}, \mathbf{e_y}$ and  wrapped on YC cylinders [see Fig.~\ref{Fig_TriangularLattice}(a)]. The system size is  $N=L_x\times L_y$, where $L_x$ ($L_y$) represents the cylinder length (circumference). The doping concentration is $\delta=N_0/N$, with $N_0$ denoting the number of doped charges, which could represent either electrons or holes, depending on the fermion charge. Here we mainly study $L_y=4,6$ and also examine $L_y=5,8$.
Due to the different rates of convergence at different parameters, the bond dimension is pushed up to $D=40,000$ when implementing $U(1)\times U(1)$ symmetry and $D=16,000$ when implementing $SU(2)\times U(1)$ symmetry.

\emph{Main findings.---} By studying the lightly doped three distinct Mott insulators (i.e., $120^{\circ}$ AF, QSL and stripe AF) with all possible hopping signs  $(\tau_1, \tau_2)$, we conclude our findings in Fig.~\ref{Fig_TriangularLattice}(c). First, the quasi-long-range superconductivity order can be realized regardless of the nature of the undoped parent Mott insulators, provided specific hopping signs. Second, the quasi-long-range superconductivity ordered phase in a triangular lattice is commonly featured by the short-ranged spin correlations and two-charge filled stripes per one-dimensional unit cell in the bulk.
Third, at larger $J_2/J_1$, sign $\tau_1$ determines the Cooper-pair phase coherence for the superconductivity phase, whereas sign $\tau_2$ is relevant to the charge pairing for all phases. Flipping the signs $\tau_1$ and $\tau_2$ from the robust superconductivity phase would give rise to pseudogap-like behavior and charge stripes without strong pairing, respectively.
Interestingly, their roles interchange at smaller  $J_2/J_1$.
Fourth, specific sign combinations $(\tau_1, \tau_2)$ would stabilize distinct phases at larger $J_2/J_1$ including SC, charge density wave (CDW), spin density wave (SDW) and pseudogap(PG)-like phase.

Unlike previous studies of the same model which only focus on $(\tau_1, \tau_2)=(-,-)$ in QSL~\cite{Jiang1, Sheng1} or $120^{\circ}$ AF~\cite{Sheng1} parent spin background to search for superconductivity,  in this work, we consider three distinct parent spin backgrounds, i.e., $120^{\circ}$ AF, QSL and stripe AF,  and all possible signs $(\tau_1, \tau_2)=(\pm,\pm)$ to understand how to achieve superconductivity, as well as revealing different correlated phases stabilized by specific sign combinations  like SC, CDW, SDW, PG-like phases. Moreover, we focus on very light doping in order to examine the role of Mott insulation, thus our focused doping concentration and $J_2/J_1$ are also different.
Considering the DMRG cylinders break the rotational symmetry, the identification
of pairing symmetry is not our focus~\cite{SM}.

\emph {Pair correlations for distinct $(\tau_1, \tau_2)$.---} We begin by examining the pair correlations 
 \begin{equation}
 D_{\alpha\beta}(\mathbf{r})\equiv \left\langle \hat{\Delta}^{\dagger}_{\alpha}(\mathbf{r_0}) \hat{\Delta}_{\beta}(\mathbf{r_0+r})\right\rangle,
 \end{equation}
where the pair operator is $\hat{\Delta}_{\alpha}(\mathbf{r}) \equiv \frac 1 {\sqrt{2}}\sum_{\sigma}\sigma {c}_ {\mathbf{r}, \sigma} {c}_ {\mathbf{r+e_\alpha}, -\sigma}$ and $\alpha,\beta=x, y, z$. 
In quasi-one-dimensional cylinders,  the true long-range superconductivity order is forbidden based on the Mermin-Wagner theorem, so we look for quasi-long-rang order  $D_{\alpha\beta}(\mathbf{r})\sim r^{-\eta_{\mathrm{sc}}}$.
In particular, $\eta_{\mathrm{sc}} < 2$ suggests divergent SC susceptibility in 2D.  

\begin{figure*}[tp]
\begin{center}
\includegraphics[width=0.9\textwidth]{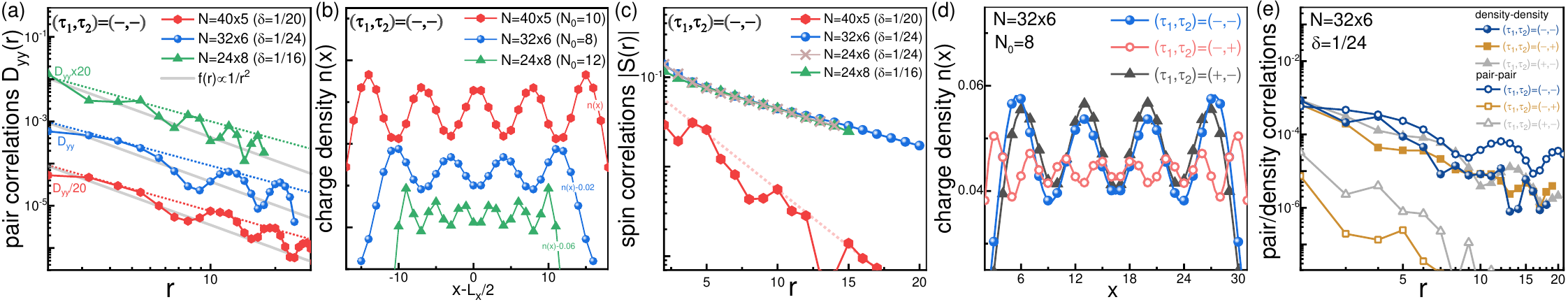}
\end{center}
\par
\renewcommand{\figurename}{Fig.}
\caption{The pair correlations (a), charge density distribution (b), spin correlations (c) on wider cylinders with $L_y=5,6,8$  for $J_2/J_1=0.24$. Starting from $(\tau_1, \tau_2)=(-,-)$ with quasi-long-range SC order, panels (d-e) show the effect of flipping the hopping sign  $\tau_2$ and $\tau_1$.
}
\label{Fig_Wide}
\end{figure*}

We compute the pair correlations for all $(\tau_1, \tau_2)$ in lightly doped spin backgrounds:  $120^{\circ}$ AF [Fig.~\ref{Fig_DD}(a,c)], QSL [Fig.~\ref{Fig_DD}(e)] and stripy AF [Fig.~\ref{Fig_DD}(g)]. We choose typical parameters of each parent phase
and find that the power-law decayed pair correlations can always be realized at certain signs $(\tau_1, \tau_2)$. As shown in Fig.~\ref{Fig_DD}(a, c), although the parent spin backgrounds are the same, the power-law-decayed pair correlations emerge at $(\tau_1, \tau_2)=(-,+)$ with $1<\eta_{\mathrm{sc}}<2$ when $J_2/J_1=0.03,0.01$ for $L_y=4$ while it switches to $(\tau_1, \tau_2)=(-,-)$ with a much slower decay rate $\eta_{\mathrm{sc}}\approx0.82$ for $J_2/J_1=0.05$.
Further increasing the ratio $J_2/J_1$ deep inside the parent spin background QSL with $J_2/J_1=0.12$ or the stripy AF with $J_2/J_1=0.24$   [see Fig.~\ref{Fig_DD}(e, g)], the pair correlations at $(\tau_1, \tau_2)=(-,-)$ become fairly strong against distance with exponent $\eta_{\mathrm{sc}}\lesssim1$, suggesting the robust superconductivity in both cases.
Moreover, the main features are qualitatively consistent when increasing the doping or the system size, as Fig.~\ref{Fig_DD}(b,d,f,h) shows, and the pair correlations become weakened with increasing doping. On wider cylinders with widths larger than $L_y=4$, as shown in Fig.~\ref{Fig_Wide} (a), we also find algebraically decayed pair correlations  at light doping with $\eta_{\mathrm{sc}}\approx 1.6$ for $L_y=5$, $\eta_{\mathrm{sc}}\approx 1.4$ for $L_y=6$ and $\eta_{\mathrm{sc}}\approx 1.3$ for $L_y=8$. The exponent $\eta_{\mathrm{sc}} < 2$ suggests robust SC with divergent susceptibility towards 2D.
These results illustrate that superconductivity can be obtained from doping distinct Mott insulators, not only from the QSL.
 More crucially, the realization of superconductivity requires specific hopping signs and finite NNN hopping.

\emph {The common features for phases with quasi-long-range superconductivity order.---}Although the parent spin backgrounds are distinct before doping, there are common features when the quasi-long-range SC order establishes after doping.

In the charge sector, we examine the charge density distribution, which is uniform along $\mathbf{e_y}$ on cylinders, so we focus on the distribution along $\mathbf{e_x}$  and define
${n}(x) \equiv 1/L_{y}\sum_{y} {\langle1-\hat{n_e}(x,y)\rangle}$.
As shown in Fig.~\ref{Fig_DD}(i) for $L_y=4$ and Fig.~\ref{Fig_Wide} (b) for $L_y=5,6,8$,  the charge profiles exhibit stripe patterns  with specific charge numbers in each stripe.
For specific $(\tau_1, \tau_2)$ with quasi-long-range SC order,
there are two doped charges per unit cell if we view  the cylinders as one-dimensional systems, i.e., $n_s=2$ with $n_s$ denoting the number of doped charges in each stripe on average, consistent with the existence of strong pairing.

In the spin sector, we compute the spin correlations $S(r)\equiv\langle \mathbf{S}({\mathbf{r_0}})\cdot\mathbf{S}({\mathbf{r_0}+r\mathbf{e_x}})\rangle$. For specific $(\tau_1, \tau_2)$ with quasi-long-range SC order, the spins commonly exhibit short-ranged correlations with finite decay length $\xi_s$,  as shown in Fig.~\ref{Fig_DD}(j) for $L_y=4$ and Fig.~\ref{Fig_Wide}(c) for wider cylinders with $L_y=5,6,8$. In particular, we find the increased decay length on wider cylinders; however, it tends to saturate when further increasing cylinder length (width) for a fixed  width (length)~\cite{SM}, e.g., unchanged $\xi_s$ when increasing $L_x$ for fixed $L_y=6$ [see Fig.~\ref{Fig_Wide}(c)]. We remark that, although the parent spin background could be distinct before doping, when the SC emerges after doping, the spin correlations become short-ranged.

\emph {The effect of hopping signs $(\tau_1, \tau_2)$ on robust superconductivity at larger $J_2/J_1$.---}  To identify the roles of hopping signs in superconductivity, we start from a robust superconductivity phase and examine the effect of flipping the signs $\tau_1$, $\tau_2$.
The robust superconductivity is characterized by power-law decayed pair correlations and the dominant pair correlations over other correlations, which occurs at $J_2/J_1\gtrsim0.05$ when $(\tau_1, \tau_2)=(-,-)$ [see Fig.~\ref{Fig_DD} (c,e,g) and Fig.~\ref{Fig_StripeSC} (a)]. To compare various correlations at light doping $\delta$, we define the renormalized correlations, including (i) the single particle propagator $\mathrm{CC(r)}\equiv [C(r)/\delta]^2$ where $C(r)=\sum_{\sigma}\langle c_{\sigma}^\dagger(\mathbf{r_0})c_{\sigma}({\mathbf{r_0}+r\mathbf{e_x}})\rangle $; (ii) the spin correlations $\mathrm{SS(r)}\equiv |S(r)|$; (iii) the pair correlations  $\mathrm{DD(r)}\equiv |D_{yy}(r)|/\delta^2$; and (iv) the charge-density correlations $\mathrm{NN(r)}\equiv |N(r)|/\delta^2$ where $N(r)=\langle n(\mathbf{r_0})n({\mathbf{r_0}+r\mathbf{e_x}})\rangle-\langle n(\mathbf{r_0})\rangle\langle n({\mathbf{r_0}+r\mathbf{e_x}})\rangle $.

As shown in Fig.~\ref{Fig_StripeSC}(a) for $(\tau_1, \tau_2)=(-,-)$, the pair correlations with exponent $\eta_{\mathrm{sc}}\approx0.96$  dominate over other correlations. We remark that, here $\eta_{\mathrm{sc}} <1<\eta_{\mathrm{cdw}}$ is consistent with the Luther-Emery liquid behavior \cite{Luther1974}.
By only switching $\tau_2$, as shown in Fig.~\ref{Fig_StripeSC}(b), the charge density correlations become dominant instead, while the pair correlations are significantly suppressed. Meanwhile, the change of $n_s$ from $2$ to $1$ [see Fig.~\ref{Fig_StripeSC}(e)] implies the breaking of Cooper pairs. This suggests the sign of NNN hopping is relevant to the charge pairing.
By contrast, when only switching $\tau_1$, we also find the suppressed pair correlations but with strong fluctuations [see Fig.~\ref{Fig_StripeSC}(c)], and all other correlations decays faster than $r^{-2}$.
Since both signs of $\tau_1$ exhibit $n_s=2$ stripes [see Fig.~\ref{Fig_StripeSC}(e)], the charge pairs are robust, consistent with a pseudogap behavior, where the doped charges form pairs but the phase coherence is lacking.
This  suggests the sign of NN hopping is relevant to the phase coherence among pairs. When simultaneously switching both signs, as shown in Fig.~\ref{Fig_StripeSC}(d), the spin correlations are remarkably enhanced and decay slightly slower than the charge-density correlations, demonstrating the robust SDWs. The spin structure factor [see the inset of Fig.~\ref{Fig_StripeSC} (d)] suggests an incommensurate SDW. We remark that the above results are obtained for doped stripe AF at $J_2/J_1=0.24$,  the doped QSL and doped $120^{\circ}$ AF at $J_2/J_1=0.05$ exhibit similar behavior~\cite{SM}.

Moreover, we further confirm the roles of $\tau_2$ and $\tau_1$ on wider cylinders. As shown in Fig.~\ref{Fig_Wide} (d) for $L_y=6$ cylinders,  when starting from $(\tau_1, \tau_2)=(-,-)$ with quasi-long-range SC order, flipping the sign $\tau_2$ would change $n_s$ from $2$ to $1$, consistent with breaking Cooper pairs. By contrast, flipping $\tau_1$ does not break Cooper pairs. However, either flipping the sign $\tau_2$ or $\tau_1$ would significantly suppress the pair correlations, as shown in Fig.~\ref{Fig_Wide} (e), while the charge-density correlations remain robust.  These observations suggest the hopping signs are relevant to charge pairing and phase coherence.

\begin{figure*}[tbp]
\begin{center}
\includegraphics[width=0.9\textwidth]{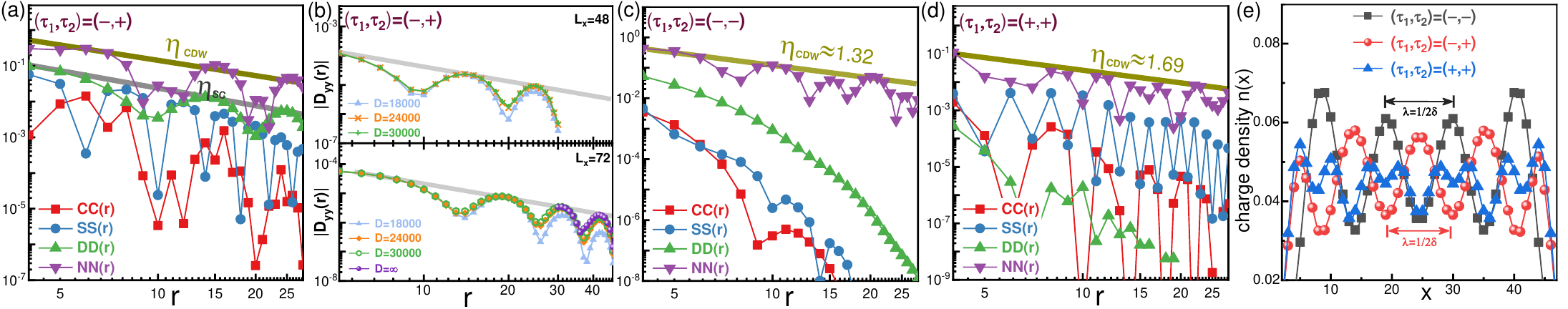}
\end{center}
\par
\renewcommand{\figurename}{Fig.}
\caption{Various correlations for distinct hopping signs $(\tau_1, \tau_2)$ at smaller $J_2/J_1$ (a,c,d). Here,  we consider $J_2/J_1=0.03$ at $\delta=1/24$ on $N=48\times4$ cylinders.
Panel (b) shows the convergence of pair correlations for different lattice sizes and panel (e) shows the charge density distributions.
Other parameters $J_2/J_1=0.02, 0.01$ exhibit similar behavior~\cite{SM}.}
\label{Fig_SmallJ2}
\end{figure*}

\emph {The effect of hopping signs $(\tau_1, \tau_2)$ on pair correlations at smaller $J_2/J_1$.---}
At smaller  $J_2/J_1$, we find algebraically decayed pair correlations  when $(\tau_1, \tau_2)=(-,+)$ with $1<\eta_{\mathrm{sc}}<2$ on $L_y=4$ cylinders [see Fig~\ref{Fig_DD}(a-b)]. Notably, the pair correlations decay at a comparable rate with the charge-density correlations, exhibiting competing quasi-long-range orders. Since $\eta_{\mathrm{cdw}}\lesssim\eta_{\mathrm{sc}}$, the stripe order is slightly dominant. The existence of competing orders is also reflected in the convergence of DMRG, which becomes much harder than larger $J_2/J_1$. With the increase of system size, it also requires a larger bond dimension to ensure the convergence of correlations at a longer distance [see Fig.~\ref{Fig_SmallJ2}(b)].
We notice that the charge profiles exhibit a two-charge filled stripe pattern only in the bulk, while the boundaries host separate single charge [see Fig.~\ref{Fig_DD}(i) and Fig.~\ref{Fig_SmallJ2}(d)].

Compared with other sign combinations, $(\tau_1, \tau_2)=(-,+)$ indeed enhances the pair correlations at smaller $J_2/J_1$ [see Fig~\ref{Fig_DD}(a)]. Now we start from $(\tau_1, \tau_2)=(-,+)$ to examine the effect of switching the signs $\tau_1$, $\tau_2$. We will show the interchanged roles of $\tau_1$, $\tau_2$ at smaller $J_2/J_1$ compared with the larger $J_2/J_1$.
After solely switching  $\tau_2$, as Fig.~\ref{Fig_SmallJ2}(b) shows,
the pair correlations are strongly suppressed. Meanwhile, the single particle propagator and spin fluctuations are also significantly suppressed.  Given the robust two-charge filled stripe pattern shown in Fig.~\ref{Fig_SmallJ2}(d), the resulting phase hosts the robust local pairing. The absence of quasi-long-range order is induced by the lack of phase coherence among pairs. Here we remark that, the pair correlations in this parameter regime also exhibit strong
anisotropy~\cite{SM}.  These observations are consistent with a pseudogap-like behavior, and the role of sign $\tau_2$ at smaller $J_2/J_1$ is similar to the role of sign $\tau_1$ at larger $J_2/J_1$ [see Fig.~\ref{Fig_StripeSC}(c)].
Furthermore, if only switching $\tau_1$, as shown in Fig.~\ref{Fig_SmallJ2}(c), the pair correlations are dramatically suppressed, while the charge profile also implies the loosened strong pairing [see Fig.~\ref{Fig_SmallJ2}(d)].
These findings suggest that $\tau_1$ may play a role in the formation of strongly paired charges, similar to the effect of $\tau_2$ at higher $J_2/J_1$[see Fig.~\ref{Fig_StripeSC}(b)].   When simultaneously switching both signs $(\tau_1, \tau_2)$, the charge density correlations become the most dominant ones \cite{SM}.

\emph {Summary and Discussion.---} In summary, we study how to realize superconductivity in lightly doped triangular-lattice Mott insulators in the strong coupling limit.
By considering three distinct undoped parent spin backgrounds and all possible sign combinations of the NN and NNN hoppings,
we find that, provided with finite NNN hopping and specific sign combinations, superconductivity can always be realized with doping distinct Mott insulators, not only the QSL. The superconductivity phase is commonly featured by short-ranged spin correlations and two-charge per stripe. Moreover, switching the sign $\tau_2$ ($\tau_1$) in a robust superconductivity phase would result in the stripe phase without strong pairing (pseudogap-like phase without Cooper-pair phase coherence) at larger $J_2/J_1$, whereas  their roles interchange at smaller  $J_2/J_1$.
We also reveal that different sign combinations stabilize distinct correlated phases, including SC, CDW, SDW and PG-like phases.
Our findings suggest the importance of kinetic energy in realizing superconductivity, which may stimulate future studies on the superconductivity mechanism and on examining different sign combinations for other lattice geometries and their intrinsic connections to previously studied square-lattice case\cite{KWu2008,ZhengZhu2018b, ShuaiChen2018,ZhengZhu2014}.
Considering different  triangular-lattice materials correspond to different sign combinations in hopping terms~\cite{NaCoO1,NaCoO2,Oike2015,Oike2017,Tang2020, Xu2020,SnSi111,Kennes2021, FengchengWu2018, Kuhlenkamp2022,Imajo2021}, our findings of distinct correlated phases stabilized by specific hopping sign combinations can be potentially probed.

\begin{acknowledgments}
Z.Z. is particularly grateful to Ashvin Vishwanath, D. N. Sheng, Shuai A. Chen for previous collaborations on the triangular-lattice Hubbard model and helpful discussions on this work. We also particularly thank insightful discussions with Zheng-Yu Weng.
This work was supported by the National Natural Science Foundation of China (Grant No.12074375), the Fundamental Research Funds for the Central Universities, the Strategic Priority Research Program of CAS (Grant No.XDB33000000) and Innovation Program for Quantum Science and Technology (Grant No. 2-6).
\end{acknowledgments}

\clearpage
\newpage

\appendix
\renewcommand{\theequation}{S\arabic{equation}}
\setcounter{equation}{0}
\renewcommand{\thefigure}{S\arabic{figure}}
\setcounter{figure}{0}
\renewcommand{\bibnumfmt}[1]{[S#1]}

\begin{widetext}
\section{Supplementary Materials for\\``{Superconductivity in doped triangular Mott insulators:
the roles of parent spin backgrounds and charge kinetic energy}"}
\end{widetext}

\begin{figure}[b!]
\begin{center}
\includegraphics[width=0.48\textwidth]{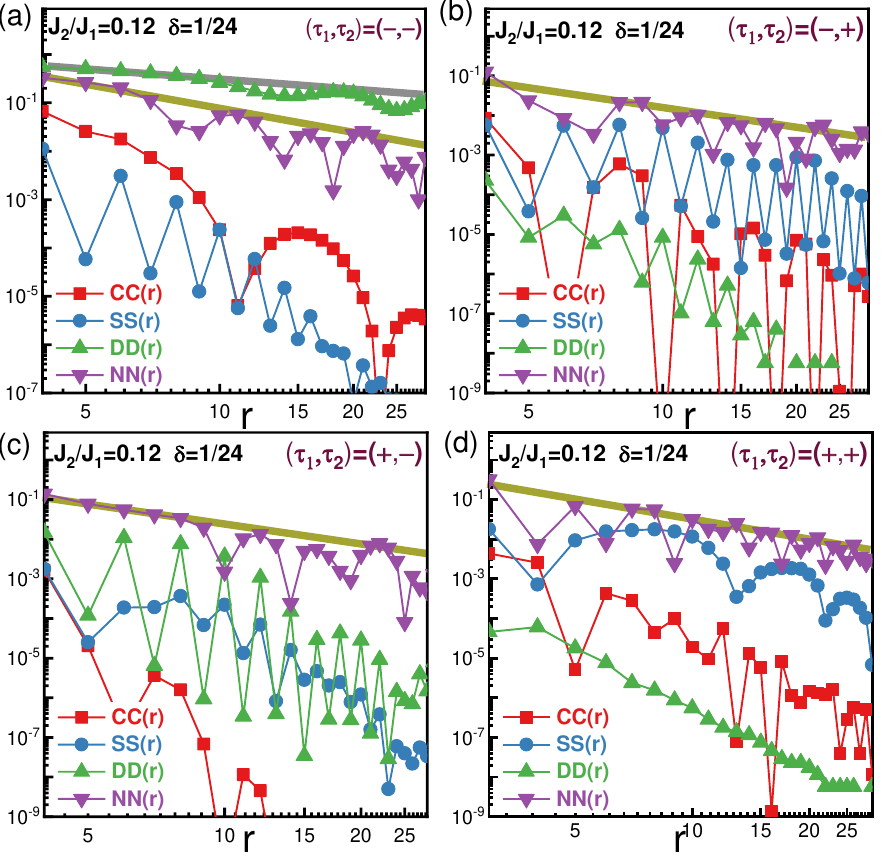}
\end{center}
\par
\renewcommand{\figurename}{Fig.}
\caption{(Color online) Various correlations versus sign combinations $(\tau_1, \tau_2)$ for $J_2/J_1=0.12$. Here, $\delta=1/24$ and $N=48\times4$.}
\label{FigS_QSL_SCSigns_124doping}
\end{figure}
\begin{figure}[b!]
\begin{center}
\includegraphics[width=0.48\textwidth]{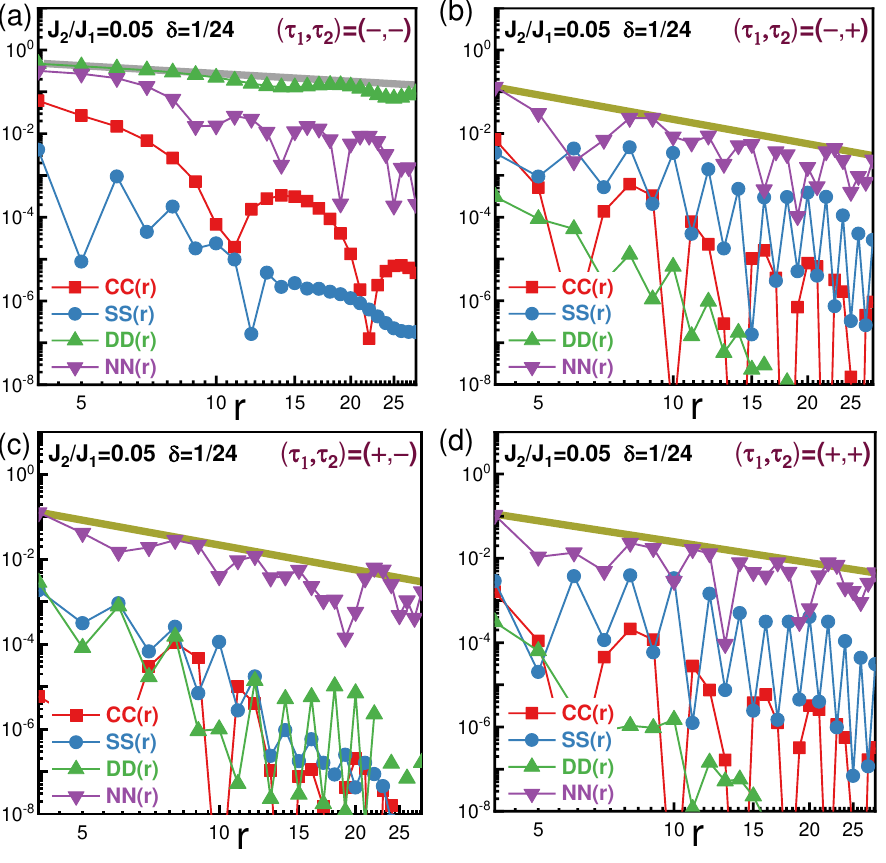}
\end{center}
\par
\renewcommand{\figurename}{Fig.}
\caption{(Color online) Various correlations versus sign combinations $(\tau_1, \tau_2)$ for $J_2/J_1=0.05$. Here, $\delta=1/24$ and $N=48\times4$.}
\label{FigS_120_SCSigns_124doping}
\end{figure}

In this work, we study the prerequisites for achieving superconductivity in the lightly doped triangular-lattice Mott insulators in the strong coupling limit. In this limit, the Hubbard model reduces to its effective Hamiltonian $t$-$J$-type models at finite doping, we consider the model including both nearest-neighbor (NN) and next-nearest-neighbor (NNN) charge hopping and spin super-exchange terms.
The Hamiltonian reads
 \begin{equation}
H=H_{\mathrm{t_1-t_2}}+H_{\mathrm{J_1-J_2}}
\end{equation}
To examine the role of Mott insulation and kinetic energy, we consider three distinct Mott insulators as the parent spin backgrounds  and  examine all possible sign combinations $(\tau_1, \tau_2)=(\pm,\pm)$ in the charge hopping terms.  The three distinct Mott insulators are controlled by the ratio of $J_2/J_1$, which are   $120^{\circ}$ antiferromagnetic (AF) at $J_2/J_1\lesssim 0.07\sim0.08$, stripy antiferromagnetic (AF) at $J_2/J_1\gtrsim0.15\sim0.16$ and the QSL in between. To examine their roles in resulting superconductivity, we focus on light doping $\delta<10\%$, particularly at $\delta=1/24$ and $\delta=1/12$, both of which show consistent results, and thus we mainly focus on specific parameters in the main text to present our findings.  In the supplementary materials, we will show more results with different parameters as the supporting data, which are consistent with the ones in the main text. These results suggest that our main findings are valid within a finite range of parameters. Numerically, we remark that the bond dimension is pushed to D=30,000 to secure the convergence of all cases, particularly considering the harder convergence at smaller ratios of $J_2/J_1$  [larger $J_2/J_1$ are easier] and specific sign combinations $(\tau_1, \tau_2)$.

\begin{figure}[tpbh]
\begin{center}
\includegraphics[width=0.49\textwidth]{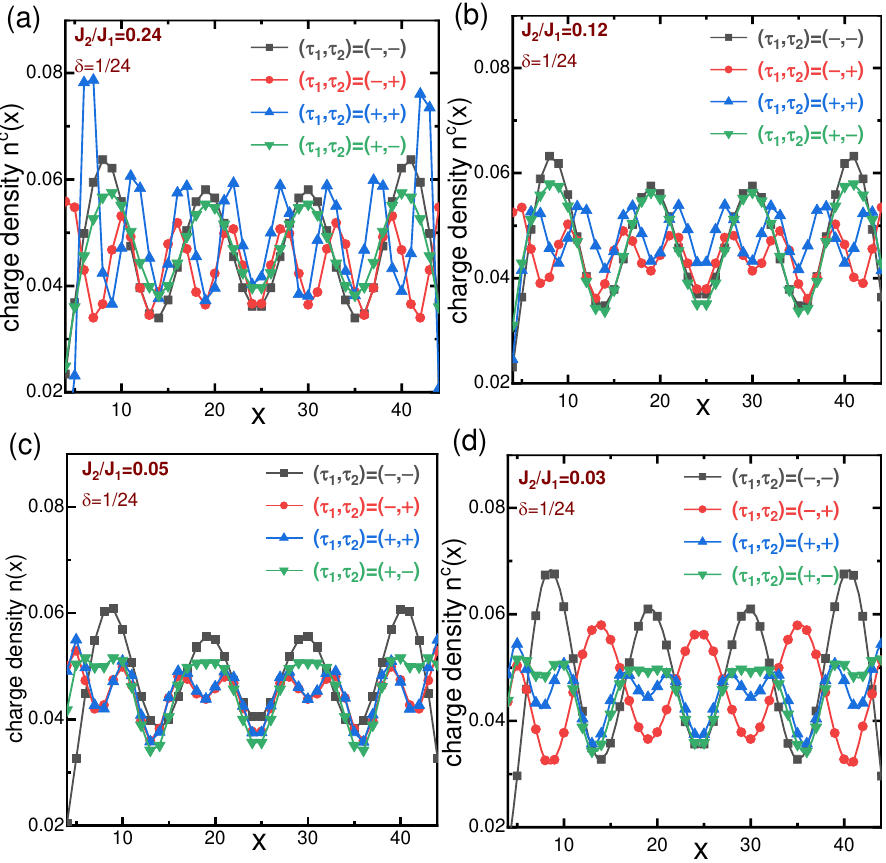}
\end{center}
\par
\renewcommand{\figurename}{Fig.}
\caption{(Color online) Charge density versus sign combinations $(\tau_1, \tau_2)$ for typical ratios of $J_2/J_1$. Here, $\delta=1/24$ and $N=48\times4$.}
\label{FigS_ChargeDensity}
\end{figure}
\begin{figure}[tpbh]
\begin{center}
\includegraphics[width=0.48\textwidth]{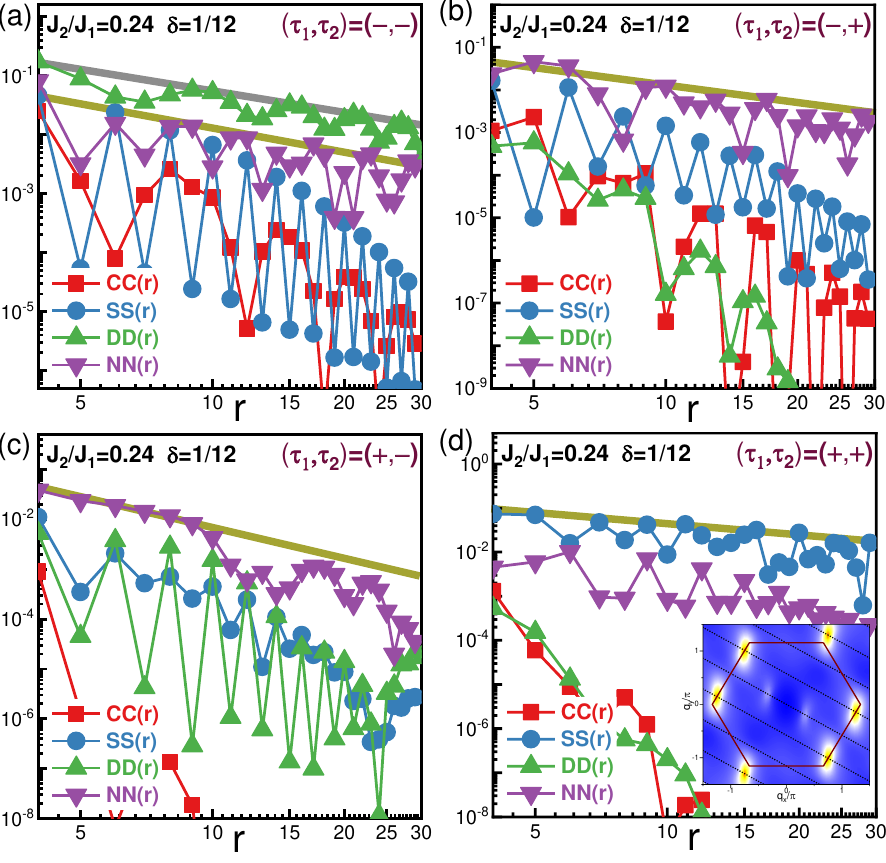}
\end{center}
\par
\renewcommand{\figurename}{Fig.}
\caption{(Color online) Various correlations versus sign combinations $(\tau_1, \tau_2)$ for $J_2/J_1=0.24$. Here, $\delta=1/12$ and $N=48\times4$.}
\label{FigS_Stripe_SCSigns_112doping}
\end{figure}
\begin{figure}[tpbh]
\begin{center}
\includegraphics[width=0.48\textwidth]{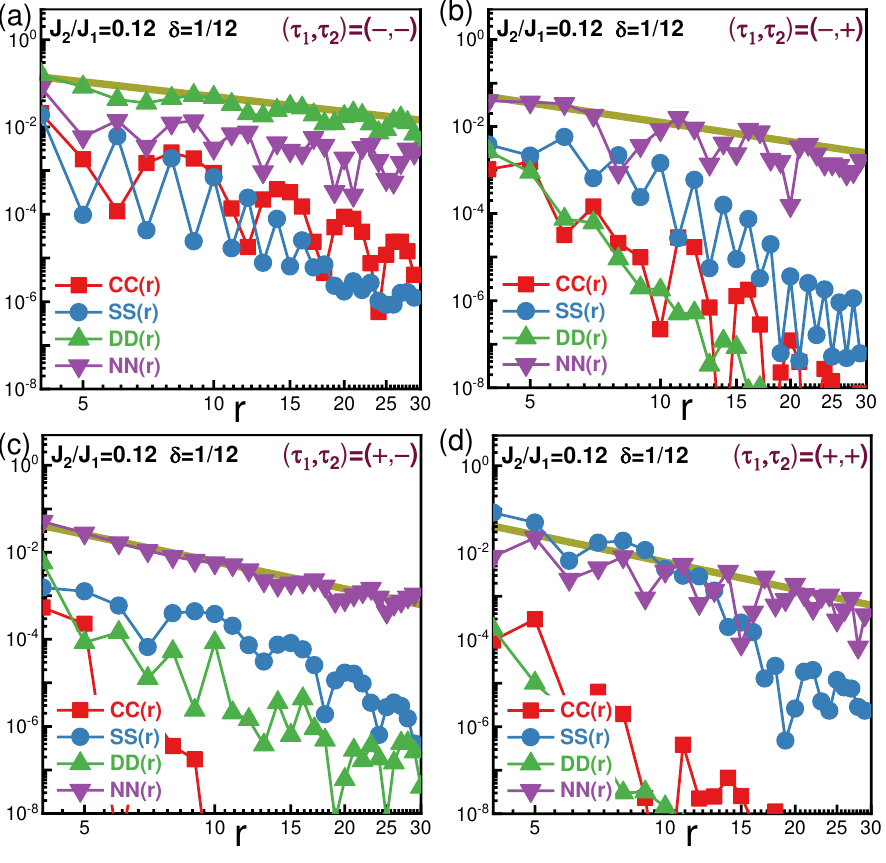}
\end{center}
\par
\renewcommand{\figurename}{Fig.}
\caption{(Color online) Various correlations versus sign combinations $(\tau_1, \tau_2)$ for $J_2/J_1=0.12$.  Here, $\delta=1/12$ and $N=48\times4$.}
\label{FigS_QSL_SCSigns_112doping}
\end{figure}
\begin{figure}[tpbh]
\begin{center}
\includegraphics[width=0.48\textwidth]{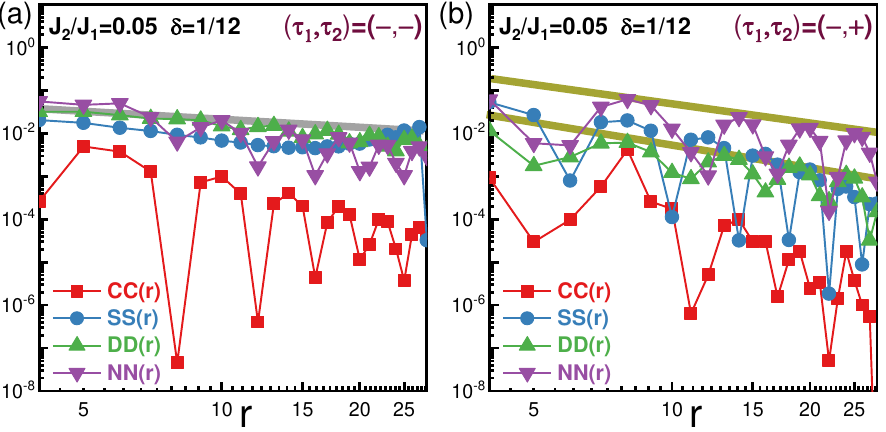}
\end{center}
\par
\renewcommand{\figurename}{Fig.}
\caption{(Color online) Various correlations versus sign combinations $(\tau_1, \tau_2)=(-,-)$ (a) and  $(\tau_1, \tau_2)=(-,+)$ (b) for $J_2/J_1=0.05$.  Here, $\delta=1/12$ and $N=48\times4$.}
\label{FigS_Critical}
\end{figure}

\begin{figure}[tpbh]
\begin{center}
\includegraphics[width=0.48\textwidth]{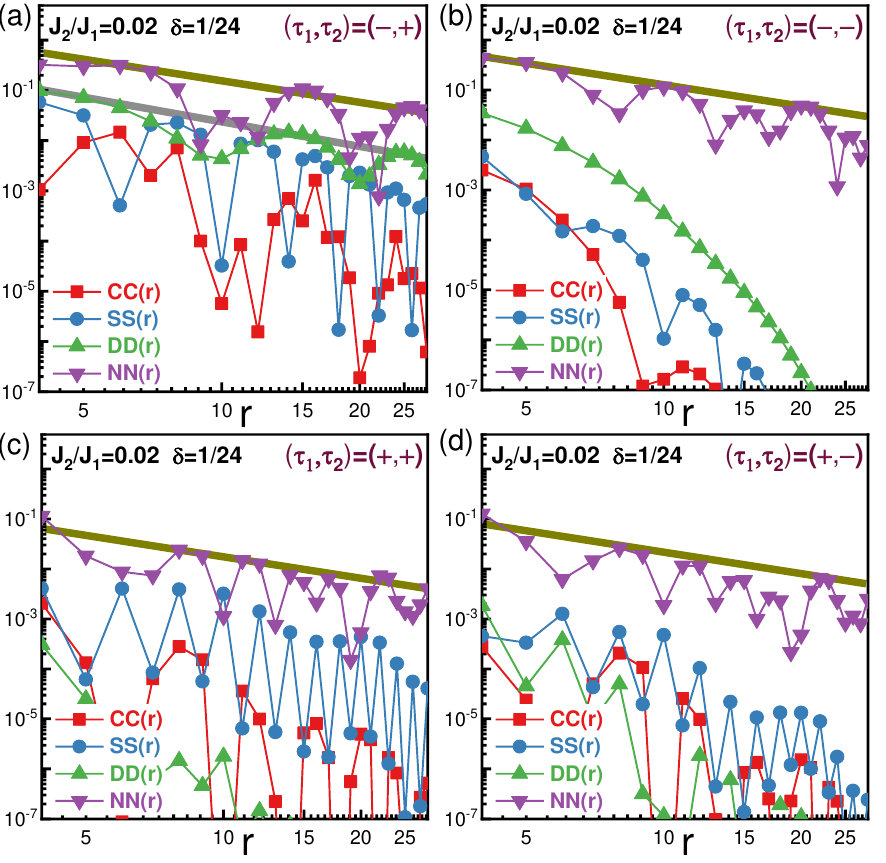}
\end{center}
\par
\renewcommand{\figurename}{Fig.}
\caption{(Color online) Various correlations versus sign combinations $(\tau_1, \tau_2)$ for $J_2/J_1=0.02$.  Here, $\delta=1/24$ and $N=48\times4$.}
\label{FigS_J2=002_124doping}
\end{figure}
\begin{figure}[tpbh]
\begin{center}
\includegraphics[width=0.48\textwidth]{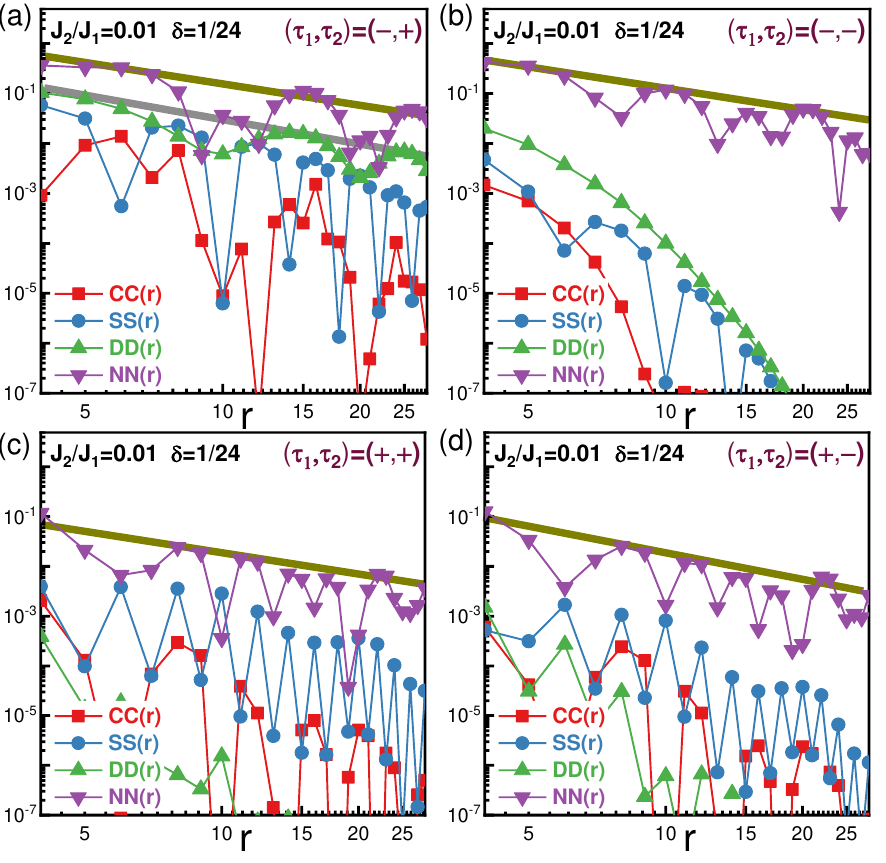}
\end{center}
\par
\renewcommand{\figurename}{Fig.}
\caption{(Color online) Various correlations versus sign combinations $(\tau_1, \tau_2)$ for $J_2/J_1=0.01$.  Here, $\delta=1/24$ and $N=48\times4$.}
\label{FigS_J2=001_124doping}
\end{figure}
\begin{figure}[tpbh]
\begin{center}
\includegraphics[width=0.48\textwidth]{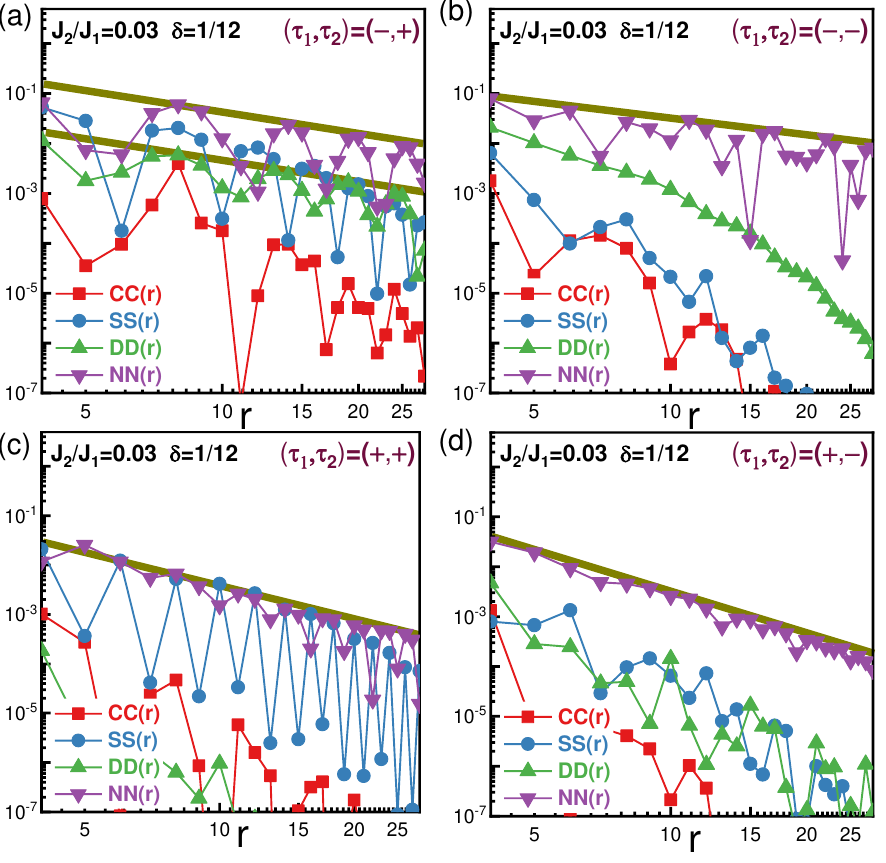}
\end{center}
\par
\renewcommand{\figurename}{Fig.}
\caption{(Color online) Various correlations versus sign combinations $(\tau_1, \tau_2)$ for $J_2/J_1=0.03$.  Here, $\delta=1/12$ and $N=48\times4$.}
\label{FigS_J2=003_112doping}
\end{figure}

\textbf{The effect of hopping signs $(\tau_1, \tau_2)$ on robust superconductivity phase at $\delta=1/24$ for doped QSL and $120^{\circ}$ AF.---}
In the main text, we have shown the results of various correlations for $J_2/J_1=0.24$ at  $\delta=1/24$ to examine the role of $\tau_1$, $\tau_2$ and their combinations $(\tau_1, \tau_2)$. $J_2/J_1=0.24$ corresponds to stripe AF spin background at half filling. Here we show that the conclusions are similar to other parent spin backgrounds, including the QSL with the ratio $J_2/J_1=0.12$ and the $120^{\circ}$ AF with the ratio $J_2/J_1=0.05$. The various correlations include
\begin{itemize}
  \item the single particle propagator $\mathrm{CC(r)}\equiv [C(r)/\delta]^2$
  \item the spin correlations $\mathrm{SS(r)}\equiv |S(r)|$
  \item the pair correlations  $\mathrm{DD(r)}\equiv |D_{yy}(r)|/\delta^2$
  \item the charge density correlations $\mathrm{NN(r)}\equiv |N(r)|/\delta^2$
\end{itemize}

Figures~\ref{FigS_QSL_SCSigns_124doping} and Figures~\ref{FigS_120_SCSigns_124doping} are results of the various correlations versus sign combinations $(\tau_1, \tau_2)=(\pm,\pm)$ for $J_2/J_1=0.12$  and $J_2/J_1=0.05$, respectively. Figures~\ref{FigS_ChargeDensity} show the charge density distribution for distinct sign combinations $(\tau_1, \tau_2)=(\pm,\pm)$.
These results are overall consistent with $J_2/J_1=0.24$ discussed in the main text. For example, the robust superconductivity, which is characterized by power-law decayed pair correlations with $\eta_{\mathrm{sc}}\lesssim1$ and the dominant pair correlations over other correlations, can be realized at signs
$(\tau_1, \tau_2)=(-,-)$. Starting from this superconductivity phase, flipping $\tau_1$ and $\tau_2$ would break the phase coherence and charge pairings, giving rise to the pseudogap-like phase and stripe phase, respectively. A slight difference arises when $(\tau_1, \tau_2)=(+,+)$,  we find the spin correlations are gradually enhanced with the increase of ratio $J_2/J_1$ and finally stabilize the SDW quasi-long range order at a larger ratio of $J_2/J_1$.
In summary, for the lightly doped three kinds of distinct Mott insulators, including stripe AF, QSL and $120^{\circ}$ AF, we find the roles of  $\tau_1$, $\tau_2$ in resulting superconductivity are consistent.

\textbf{The effect of hopping signs $(\tau_1, \tau_2)$ on robust superconductivity phase for larger ratio $J_2/J_1$ at other dopings
---}
In the main text, we mainly show the results at doping $\delta=1/24$ to examine the roles of $\tau_1$, $\tau_2$ and their combinations $(\tau_1, \tau_2)$.  Here we present the results at a different doping concentration $\delta=1/12$ and confirm that the conclusions in the main text are valid at a different doping concentration.
As shown in Fig.~\ref{FigS_Stripe_SCSigns_112doping} for $J_2/J_1=0.24$ and Fig.~\ref{FigS_QSL_SCSigns_112doping} for $J_2/J_1=0.12$, the results at   $\delta=1/12$ are consistent with the ones with the same $(\tau_1, \tau_2)$ at $\delta=1/24$. Only the incommensurate momenta of SDW at $(\tau_1, \tau_2)=(+,+)$  slightly shift with increasing doping for $J_2/J_1=0.24$. In summary, these results suggest that our main findings are valid within a finite range of doping.

Our main findings are summarized in Fig.1(c) in the main text, which is based on $\delta=1/24$ doping. In the above, we have shown that the main findings are robust against a finite doping range.
If we fix the sign combinations $(\tau_1, \tau_2)$ and increase the ratio $J_2/J_1$, we can find there is a phase transition between $J_2/J_1=0.03$ and $J_2/J_1=0.05$. When the ratio $J_2/J_1$ is smaller than this critical ratio, the enhanced pair correlations appear at $(\tau_1, \tau_2)=(-,+)$, while when the ratio $J_2/J_1$ is larger than this critical ratio, the robust superconductivity phase appears at $(\tau_1, \tau_2)=(-,-)$. Here, we remark that, as one can expect, such a transition point would slightly shift  with doping. For example, when we study the doping $\delta=1/12$, we find that such a critical ratio shifts to a larger ratio of $J_2/J_1\sim0.05$. As shown in Fig.~\ref{FigS_Critical} (a-b), both $(\tau_1, \tau_2)=(-,+)$ and $(\tau_1, \tau_2)=(-,-)$ show enhanced pair correlations.  For the doping $\delta=1/12$, as we have shown in Fig.~\ref{FigS_Stripe_SCSigns_112doping} for $J_2/J_1=0.24$ and Fig.~\ref{FigS_QSL_SCSigns_112doping} for $J_2/J_1=0.12$, the robust superconductivity quasi-long-range order sets up at $(\tau_1, \tau_2)=(-,-)$ for larger $J_2/J_1$. By contrast, the enhanced pair correlations only appear at $(\tau_1, \tau_2)=(-,+)$ for $J_2/J_1\leqslant0.03$, as shown in Fig.~\ref{FigS_J2=003_112doping}. Then one can expect a transition around the ratios $J_2/J_1$ in between, and the behavior of $J_2/J_1=0.05$ [see Fig.~\ref{FigS_Critical}]  is close to such a critical regime when doping $\delta=1/12$.

\begin{figure}[tpbh]
\begin{center}
\includegraphics[width=0.48\textwidth]{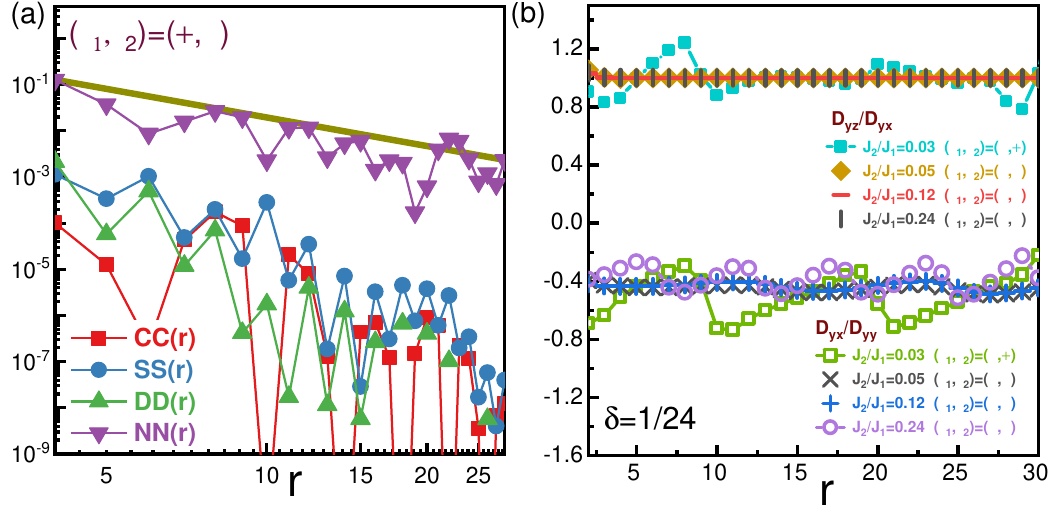}
\end{center}
\par
\renewcommand{\figurename}{Fig.}
\caption{(Color online) (a) Various correlations for $(\tau_1, \tau_2)=(+,-)$. Here,  we consider $J_2/J_1=0.03$ at $\delta=1/24$ on $N=48\times4$ cylinders. (b) The ratio of  pair correlations $D_{\alpha\beta}$ between different directional bonds.  Here, $\delta=1/24$ and $N=48\times4$.}\label{FigS_ZhengFu_correlations_Pair_ratio}
\end{figure}

\textbf{The effect of hopping signs $(\tau_1, \tau_2)$ on enhanced pair correlations for smaller ratio $J_2/J_1$
 ---}
In the main text, we mainly show the results of smaller ratios of $J_2/J_1$ at $\delta=1/24$ doping. In Fig.~\ref{FigS_ZhengFu_correlations_Pair_ratio}(a), we show the resulting phase when simultaneously flipping the sign $(\tau_1, \tau_2)$, the pair correlations are strongly suppressed while only leaving the charge density correlations as the most dominant correlations, suggesting a robust charge stripe phase. In Fig.~\ref{FigS_J2=002_124doping}and Fig.~\ref{FigS_J2=001_124doping}, we present the results for other values of $J_2/J_1$, including $J_2/J_1=0.02$ [see Fig.~\ref{FigS_J2=002_124doping}] and $J_2/J_1=0.01$ [see Fig.~\ref{FigS_J2=001_124doping}], they all show consistent numerical observations with $J_2/J_1=0.03$ shown in the main text.
Moreover, here we supplement the results for the same $J_2/J_1$  mainly at doping concentration $\delta=1/12$. As shown in Fig.~\ref{FigS_J2=003_112doping}, we find the results for $\delta=1/12$ are consistent with the main text. So we can find the main findings of smaller $J_2/J_1$ in the main text are also valid at a finite range of ratios and doping concentrations.

\begin{figure}[tpb]
\begin{center}
\includegraphics[width=0.48\textwidth]{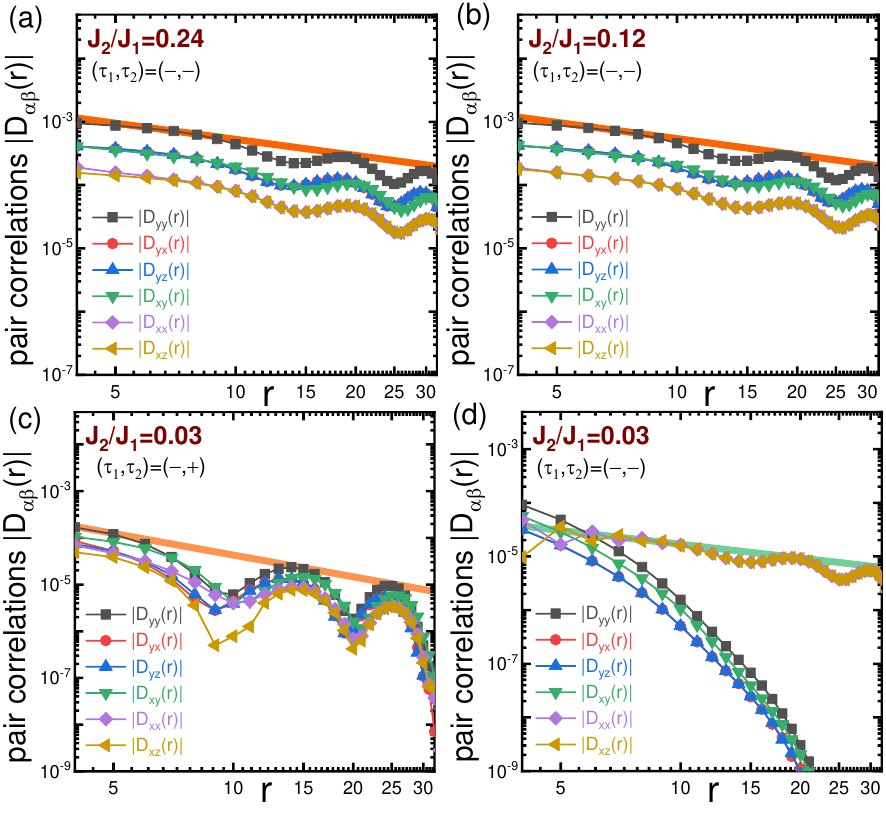}
\end{center}
\par
\renewcommand{\figurename}{Fig.}
\caption{(Color online) The pair correlations between bonds along $\alpha,\beta=x,y,z$ directions in doped distinct spin backgrounds with specific sign combinations. Here, $\delta=1/24$ and $N=48\times4$.}\label{FigS_AnisotropicSC}
\end{figure}

\begin{figure}[tpbh]
\begin{center}
\includegraphics[width=0.48\textwidth]{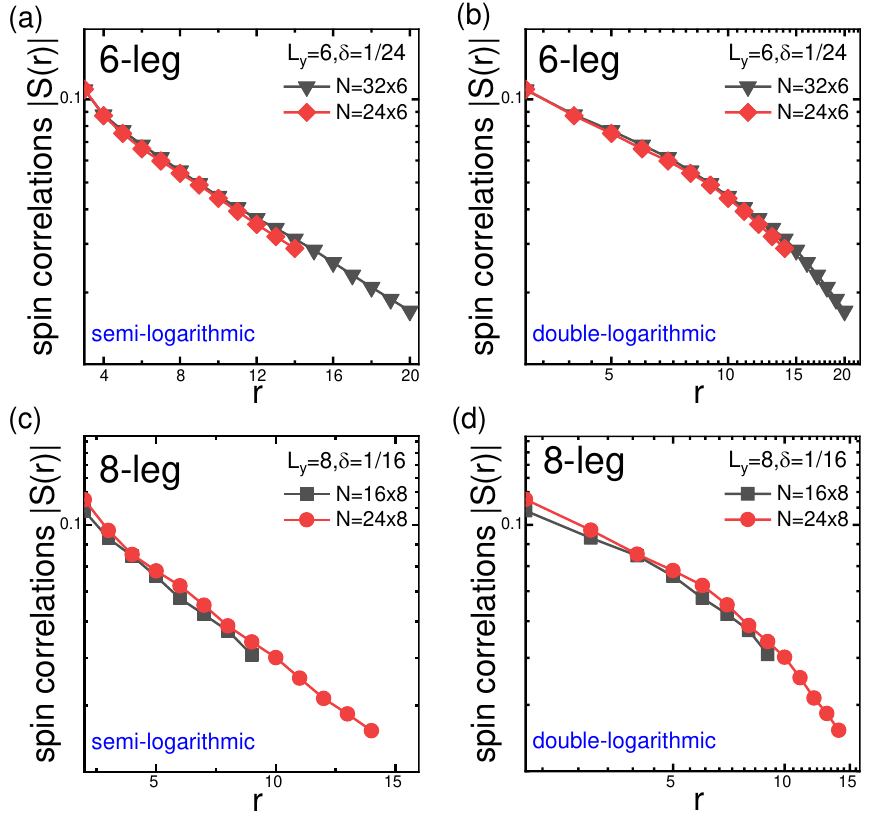}
\end{center}
\par
\renewcommand{\figurename}{Fig.}
\caption{(Color online) The spin correlations on $L_y=6$ (a-b) and $L_y=8$ cylinders (c-d). The left panels in each row show the plots in a semi-logarithmic scale (a, c), while the right panels are the plots of the same data but in a double-logarithmic scale (b,d).
Here, $J_2/J_1=0.24$ }\label{FigS_SpinCor}
\end{figure}

\begin{figure}[tpb]
\begin{center}
\includegraphics[width=0.48\textwidth]{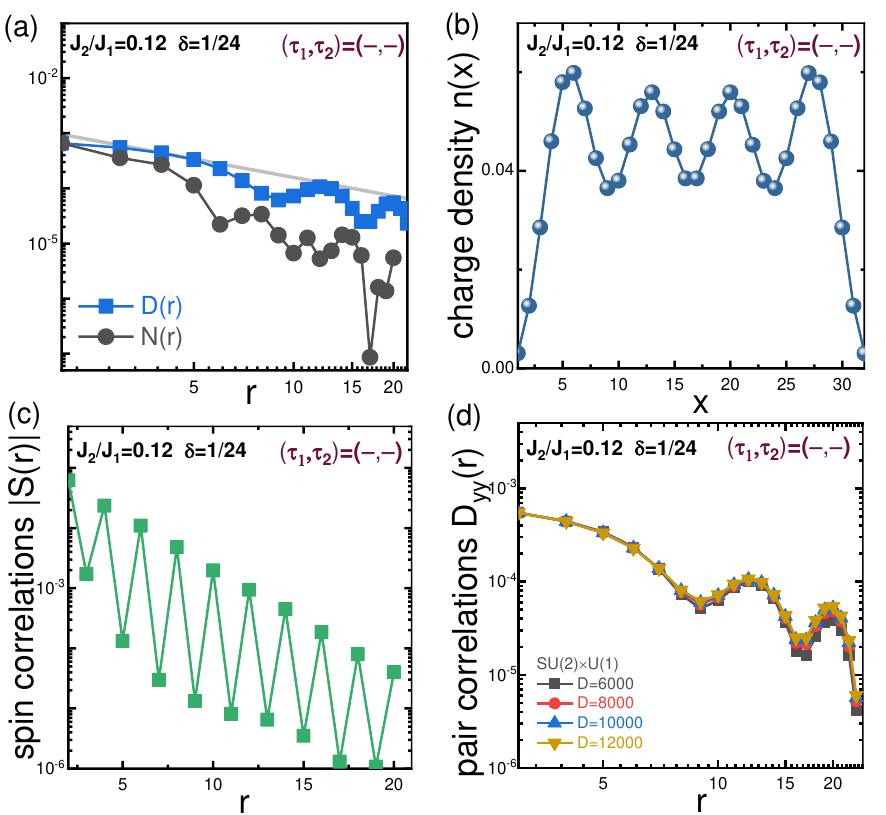}
\end{center}
\par
\renewcommand{\figurename}{Fig.}
\caption{(Color online) (a) the pair correlations and charge-density correlations, (b) the charge density distributions, and (c) the spin-spin correlations for $L_y=6$ cylinders. (d) The bond dimension dependence of the pair correlations. the Here, $J_2/J_1=0.12$  and $\delta=1/24$ on $N=32\times6$ cylinder.  }\label{FigS_FigS_6legQSL}
\end{figure}

\textbf{The pair correlations among different types of bonds ---}
In the main text, we mainly focus on how to achieve superconductivity by examining distinct parent spin backgrounds and all possible combinations of charge hopping signs. The detailed pairing symmetry is not our main focus, particularly considering the DMRG cylinders break the spatially rotational symmetry in nature.
Because of the rotational symmetry breaking on cylinders,
the specific pairing symmetry might also  be lifted. Here, we present our numerical observations on the pair correlations among different directional bonds.

As shown in Fig.~\ref{FigS_ZhengFu_correlations_Pair_ratio}(b), we find the ratio between $D_{yx}$ and $D_{yz}$ are close to 1, suggesting the equal amplitude and signs for the pair correlations between these two types of bonds. By  contrast, $D_{yx}/D_{yy}$ are negative and close to $-0.4$, indicating the different signs between these two types of correlations, and the smaller amplitude for the pair correlations between $y$ and $x$ bonds. These features are consistent with d-wave-like pairing. We remark that it is hard to distinguish whether  the nematic nature of pairing is intrinsic or just due to the lattice geometry effect. For the smaller ratio of $J_2/J_1$ or larger ratio of $J_2/J_1$, the profile exhibits additional oscillations,  while they become uniform for the intermediate ratios of $J_2/J_1$.

In Fig.~\ref{FigS_AnisotropicSC}, we show the pair correlations of $D_{\alpha\beta}$ versus distance for distinct $\alpha,\beta$. In the robust superconductivity phase with  $(\tau_1, \tau_2)=(-,-)$ at larger $J_2/J_1$ [see Fig.~\ref{FigS_AnisotropicSC} (a-b)] or the enhanced pair correlations with $(\tau_1, \tau_2)=(-,+)$ at smaller $J_2/J_1$ [see Fig.~\ref{FigS_AnisotropicSC} (c)], different types of pair correlations exhibit consistent behavior, and we mainly focus on $D_{yy}$ in the main text. Here we would like to point out that, in the pseudogap-like phase at smaller $J_2/J_1$ with $(\tau_1, \tau_2)=(-,-)$, we find the rapid decay of $D_{yy}$ but the quasi-long-range ordered pair correlations for $D_{xx}$. Such anisotropic behavior suggests the absence of uniform quasi-long-ranged superconductivity orders, consistent with pseudogap-like behavior, particularly considering the rapidly decayed single particle propagator and the charge pairs formed in this parameter regime.

\textbf{The spin correlations on wider cylinders. ---} In the main text, we have discussed that, for specific $(\tau_1, \tau_2)$ with quasi-long-range SC order, the spins commonly exhibit short-ranged correlations with finite decay length  $\xi_s$. In particular, we notice that, compared with the narrower cylinders (i.e., $L_y=4,5$), the decay length of spin correlations increases on wider cylinders ($L_y=6,8$). In Fig.~\ref{FigS_SpinCor}, we plot the same correlations on  $L_y=6$ cylinders [see Fig.~\ref{FigS_SpinCor}(a-b)] and $L_y=8$ cylinders [see Fig.~\ref{FigS_SpinCor}(c-d)] in both semi-logarithmic scale and double-logarithmic scale, which suggests the spin correlations still decay exponentially because the straight line is better exhibited in semi-logarithmic scale. Moreover, for a fixed width $L_y$, when we increase the cylinder length $L_x$, the decay length of the spin correlations remain the same [see Fig.~\ref{FigS_SpinCor}(a-b) or Fig.~\ref{FigS_SpinCor}(c-d)], implying the saturated decay length on $L_y=6,8$ cylinders, though they are larger than narrow cylinders  with $L_y=4,5$.

\textbf{The additional results on wider cylinders. ---}In the main text, we have shown the quasi-long-range SC order at  $J_2/J_1=0.24$. In Fig.~\ref{FigS_FigS_6legQSL}, we also compute the lightly doped QSL parent spin background at $J_2/J_1=0.12$.  Here, we consider $N=32\times6$ cylinder at  $\delta=1/24$ doping. Fig. ~\ref{FigS_FigS_6legQSL} (a) shows both the pair correlations and charge-density correlations decay algebraically with $1<\eta_{\mathrm{SC}}<2$, $1<\eta_{\mathrm{CDW}} <2$  and with $\eta_{\mathrm{SC}}<\eta_{\mathrm{CDW}}$. Since $\delta=1/24$ doping corresponds to the doped charge number $N_0=8$, as Fig. ~\ref{FigS_FigS_6legQSL} (b) shows, there are $N_0/2$ peaks in the charge distribution along cylinders ( i.e., along $\mathbf{e_x}$), consistent with the existence of strong pairing. Meanwhile, the spin correlations decay exponentially [see the semi-logarithmic plot in Fig. ~\ref{FigS_FigS_6legQSL} (c)]. In  Fig. ~\ref{FigS_FigS_6legQSL} (d), we show the bond dimension dependence of the pair correlations at $J_2/J_1=0.12$, demonstrating the good convergence of the results without bond dimension scaling of the pair correlations in this case.

\end{document}